\documentclass[conference]{IEEEtran}
\IEEEoverridecommandlockouts
\IEEEpubid{\makebox[\columnwidth]{DOI: 10.1109/DASC.2017.8102145~\copyright~2017 IEEE\hfill} \hspace{\columnsep}\makebox[\columnwidth]{ }}
\usepackage[english]{babel}
\usepackage{calc} 
\usepackage{scrextend}
\usepackage[single]{acro}
\usepackage{booktabs, caption, tabularx, makecell}
\usepackage{cite}
\usepackage{comment}
\usepackage{array,ragged2e}
\DeclareAcronym{soc}{short = SoC, long = System-on-a-Chip, foreign-lang = english,	single-format = \MakeLowercase}

\DeclareAcronym{ea}{
	short = E/A,
	long  = Eingabe/Ausgabe ,
}

\DeclareAcronym{hwcca}{
	short = HWCCA,
	long = Hardware COTS Component Assurance,
	long-plural-form  = Hardware COTS Component Assurance,
	foreign-lang = english,
	single-format = \MakeLowercase
}

\DeclareAcronym{da}{
	short = DA,
	long = Design Assurance,
	long-plural-form  = Design Assurance,
	foreign-lang = english,
	single-format = \MakeLowercase
}

\DeclareAcronym{dan}{
	short = DAN,
	long = Development Assurance Niveau,
	long-plural-form  = Development Assurance Niveau,
	foreign-lang = english,
	single-format = \MakeLowercase
}

\DeclareAcronym{seh}{
	short = SEH,
	long = Simple Electronic Hardware,
	long-plural-form  = Simple Electronic Hardware,
	foreign-lang = english,
	single-format = \MakeLowercase
}

\DeclareAcronym{ceh}{
	short = CEH,
	long = Complex Electronic Hardware,
	long-plural-form  = Complex Electronic Hardware,
	foreign-lang = english,
	single-format = \MakeLowercase
}

\DeclareAcronym{smcu}{
	short = SMCU,
long = Safety Microcontroller Unit,
long-plural-form  = Safety Microcontroller Units,
foreign-lang = english,
single-format = \MakeLowercase
}

\DeclareAcronym{cri}{
	short = CRI,
	long  = Certification Review Item ,
}

\DeclareAcronym{ip}{
	short = IP,
	long  = Intellectual Property,
	long-plural-form  = Intellectual Properties,
}

\DeclareAcronym{ee}{
	short = e/e,
	short-plural-form = e/e,
	long  = elektrische/elektronische,
	long-plural-form = elektrischen/elektronischen,
	single-format = \MakeLowercase
}

\DeclareAcronym{uav}{
	short = UAV,
	long  = Unmanned Air Vehicle,
	single-format = \MakeLowercase
}

\DeclareAcronym{amc}{
	short = AMC,
	long  = Acceptable Means of Compliance,
	single-format = \MakeLowercase
}
\DeclareAcronym{aeh}{
	short = AEH,
	long  = Airborne Electronic Hardware,
	long-plural-form = Airborne Electronic Hardware,
	short-plural-form = AEH,
	single-format = \MakeLowercase
}
\DeclareAcronym{fc}{
	short = FC,
	long  = failure condition
}
\DeclareAcronym{easa}{
	short = EASA,
	long  = European Aviation Safety Agency,
	single-format = \MakeLowercase
}
\DeclareAcronym{cs}{
	short = CS,
	long  = Certification Specification,
		single-format = \MakeLowercase
}

\DeclareAcronym{mcu}{
	short = MCU,
	long = Microcontroller Unit,
	foreign-lang = english,
		single-format = \MakeLowercase
}

\DeclareAcronym{asic}{
	short = ASIC,
	long = anwendungsspezifische integrierte Schaltung,
	foreign  = Application-Specific Integrated Circuit,
	long-plural = en,
	foreign-lang = english
}

\DeclareAcronym{amcu}{
	short = AMCU,
	long = Automotive Microcontroller Unit,
	foreign-lang = english,
		single-format = \MakeLowercase
}

\DeclareAcronym{mpu}{
	short = MPU,
	long = Microprocessor Unit,
		single-format = \MakeLowercase
}

\DeclareAcronym{wcet}{
	short = WCET,
	long  = Worst Case Execution Time,
	single-format = \MakeLowercase
}

\DeclareAcronym{ccdl}{
	short = CCDL,
	long  = Cross Channel Data Link,
	single-format = \MakeLowercase
}

\DeclareAcronym{cots}{
	short = COTS,
	long  = Commercial off-the-shelf,
	single-format = \MakeLowercase
}
\DeclareAcronym{dmac}{
	short = DMA-C,
	long  = Direct Memory Access Controller,
	single-format = \MakeLowercase
}

\DeclareAcronym{mmu}{
	short = MMU,
	long  = Memory Management Unit,
	single-format = \MakeLowercase}
\DeclareAcronym{ima}{
	short = IMA,
	long  = Integrated Modular Avionic,
	single-format = \MakeLowercase
}

\DeclareAcronym{ecmp}{
  short = ECMP,
  long  = Electronic Component Management Plan,
    single-format = \MakeLowercase
}

\DeclareAcronym{seooc}{
	short = SEooC,
	long  = Safety Element out of Context,
	single-format = \MakeLowercase
}
\DeclareAcronym{swap}{
	short = SWaP,
	long  = {Space, Weight and Power},
	single-format = \MakeLowercase
}

\DeclareAcronym{hwdal}{
	short = HWDAL,
	long  = Hardware Design Assurance Level,
	single-format = \MakeLowercase
}

\DeclareAcronym{dal}{
	short = DAL,
	long  = Development Assurance Level,
	single-format = \MakeLowercase
}

\DeclareAcronym{pse}{
	short = PSE,
	long  = Product Service Experience,
	single-format = \MakeLowercase
}

\DeclareAcronym{lru}{
	short = LRU,
	long  = Line Replaceable Unit,
	single-format = \MakeLowercase
}

\DeclareAcronym{asil}{
	short = ASIL,
	long  = Automotive Safety Integrity Level,
	single-format = \MakeLowercase
}

\DeclareAcronym{fcc}{
	short = FCC,
	long  = Flight Control Computer,
	single-format = \MakeLowercase
}

\DeclareAcronym{bist}{
	short = BIST,
	long  = Build In Self Test,
	single-format = \MakeLowercase
}

\DeclareAcronym{aadl}{
	short = AADL,
	long  = Architecture Analysis and Design Language
}

\DeclareAcronym{faa}{
	short = FAA,
	long  = Federal Aviation Administration,
	single-format = \MakeLowercase
}

\DeclareAcronym{seu}{
	short = SEU,
	long  = Single Event Upset,
	single-format = \MakeLowercase
}

\DeclareAcronym{ca}{
	short = CA,
	long  = Certification Authority,
	single-format = \MakeLowercase
}

\begin{document}


\title{Design Assurance Evaluation of Microcontrollers for safety critical Avionics}

\author{
\IEEEauthorblockN{Andreas Schwierz}
\IEEEauthorblockA{Research Center:\\Competence Field Aviation\\ Technische Hochschule Ingolstadt\\ 85049 Ingolstadt, Germany\\ Email: Andreas.Schwierz@thi.de}
\and
\IEEEauthorblockN{H\r{a}kan Forsberg}
\IEEEauthorblockA{School of Innovation, Design and Engineering\\Division of Intelligent Future Technologies\\ M\"alardalen University\\ 721 23 V\"aster\r{a}s, Sweden\\ Email: hakan.forsberg@mdh.se}
}


\maketitle

	\begin{abstract}
Dealing with \ac{cots} components is a daily business for avionic system manufacturers. They are necessary ingredients for hardware designs, but are not built in accordance with the avionics consensus standard DO-254 for \ac{aeh} design. Especially for complex \ac{cots} hardware components used in safety critical \ac{aeh}, like \acp{mcu}, additional assurance activities have to be performed. All of them together shall form a convincing confident, that the hardware is safe in its intended operation environment. The focus of DO-254 is one approach called \emph{\ac{da}}. Its aim is to reduce design errors by adherence of prescribed process objectives for the entire design life cycle. The effort for certain \ac{cots} assurance activities could be reduced if it is possible to demonstrate, that the \ac{cots} design process is based on similar effective design process guidelines to minimize desgin errors. In the last years, semiconductor manufacturers released safety \acp{mcu} in compliance to the ISO 26262 standard, dedicated for the development of functional safe automotive systems. These products are \ac{cots} components in the sense of avionics, but they are also developed according to a process that focuses on reduction of design errors. In this paper an evaluation is performed to figure out if the ISO 26262 prescribes a similar \ac{da} approach as the DO-254, in order to reduce the \ac{cots} assurance effort for coming avionic systems.
\end{abstract}
	


\section{Introduction}\label{sec:intro}

For development of \ac{aeh} \ac{cots} components are an inescapable ingredient. From \ac{seh} such as serial peripheral controller, to \ac{ceh}, like a \ac{mpu}, \ac{cots} components are widely used in safety critical avionic systems today. The availability of suitable \acp{mpu} for safety critical \ac{aeh} is reducing. \acp{mcu}\footnote{Here also referenced as \ac{cots} component.} have a bigger market share and should be considered for future developments. Their high degree of functional integration can realize the majority of avionic system design requirements. However, the inherent complexity of a \ac{soc} is very high and methods for \ac{cots} assurance to establish certification conformity are still an open topic of discussion in aerospace~\cite{AFE75ProjectManagementCommittee.}. The reason is that, function blocks formerly developed in own responsibility by aviation industry are now part of the \ac{mcu}. Its whole design life cycle became supplier controlled and is not necessarily aligned to \ac{aeh} needs coming from certification or targeted operation context. The certification applicant has to demonstrate that the \ac{cots} component reached the same level of confidence as it would be achievable by an in-house design using aviation processes, w.r.t. the reasonably reduction of design errors. The concept of this process based evidence is called \ac{da}\footnote{In context of airborne system and software development also known as development assurance.}. It claims a structured and requirement driven process intended to mitigate design errors, which could not be exposed otherwise due to the potential lack of complete verification~\cite{FAA.2001}. The industry consensus standard DO-254~\cite{RTCA.19.4.20001} and further guidance of certification authorities define the \ac{da} approach for \ac{aeh} and how it could be adapted depending on different levels of safety implications induced by the hardware in a given system context. In order to be aligned with certification recommendations, adherence of that approach is an integral part of the argumentation about safe operation of an \ac{aeh} or its inherent \ac{cots} components. 

Summarized, \acp{mcu} shall be integrated certification compliant  into the \ac{aeh} and it is necessary within an \ac{cots} assurance process to demonstrate that the \ac{mcu} design process produced the level of integrity\footnote{Convincing confident about no unexposed design errors reaching the final product.} required for \ac{aeh}. An evaluation between an appropriate \ac{mcu} design process and the DO-254 is needed to determine it.



This paper is organized as follows:
In section \ref{sec:certPrac} a short overview  is given about current industrial practises for \ac{cots} assurance and research projects that contributed in this field. Section \ref{sec:DAE} describes in \ref{sec:Challenge} and \ref{sec:Method} the challenge and method for the evaluation between the DO-254 and an \ac{mcu} design process. In section \ref{sec:Realisation} the evaluation is performed and findings for each examined DO-254 hardware design life cycle process are delivered. Section \ref{sec:Conclusion} contains a conclusion with ideas for future work.


%

\section{Certification Practice}\label{sec:certPrac}

Although the \ac{faa} and the \ac{easa} have different positions about the application scope~\cite{.f}, ~\cite{easa.2016} of the guidance document DO-254~\cite{RTCA.19.4.20001}, it is still the current recommended industry consensus standard for certification compliant development of \ac{aeh}. DO-254 deals with a process model that is created around the principle of design assurance and contains additional considerations for the use of \ac{cots} components, mainly through an electronic component management process (section 11.2 in~\cite{RTCA.19.4.20001}) and product service experience (section 11.3 in~\cite{RTCA.19.4.20001}). 
In the past, it was unsuccessfully attempted to demonstrate DO-254 design assurance compliance by means of reverse engineering of required process artefacts of an \ac{mpu}. After years, it was clear that this is not a practical approach~\cite{Hilderman.20073}. With this lack in mind the certification authorities assigned several research projects to investigate the potential risks and challenges of different \ac{cots} technologies and how they can be addressed to satisfy certification requirements. As a result, further guidance materials were derived to enable complex \ac{cots} component usage in \ac{aeh}.

Between 2006 and 2011, The Aerospace Vehicle Systems Institute (AVSI) project Authority for Expenditure (AFE) No. 43, released five research reports (\cite{Dr.RabiMahapatra(TexasA&MUniversity)withinAVSICooperativeoversight.12.2006}, \cite{NTIS.02.2009}, \cite{AVSI.06.2008}, \cite{NTIS.05.2011}, \cite{AIR120.09.2010}) and a handbook~\cite{AIR100.02.2011} addressing the use of \acp{mpu} in airborne systems. These reports and the handbook introduced the concept of safety nets, i.e. the employment of mitigations and protections at appropriate levels to protect against unexpected behaviour of an \ac{mpu}. In the following AVSI AFE-75 project (performed between 2011-2013), the scope was extended to cover all kinds of \ac{cots} components including \ac{cots} assemblies. AVSI AFE-75 found more than 20 issues on the use of \ac{cots} electronics in \ac{aeh}~\cite{AFE75ProjectManagementCommittee.}. Special attention was given to attain a consistent application of safety and reliability guidance for all aerospace companies certifying under \ac{faa}\cite{Mutuel.2016}.
In 2012, \ac{easa} released the certification memorandum (CM) CM-SWCEH-001 \cite{EASA.9.5.20121} to address design assurance of airborne electronic hardware. Section 9 in the CM-SWCEH-001 addresses guidelines for \ac{cots} digital \ac{aeh} while section 10 covers the use of \ac{cots} graphical processors. Through section 9, \ac{easa} extends the application scope of DO-254 to include complex and even highly complex \ac{cots} components such as multicore processors as part of the \ac{aeh} design. This is performed through different activities (from 2 up to 16), depending on device complexity and design assurance level. These activities were partially derived by research findings in~\cite{FAUBLADIER.20081}. The refined process description is a reasonable guidance coming from experience over many certification projects. In addition, the CM allows using alternative methods for \ac{cots} devices, such as reverse engineering of DO-254 life cycle data or other means of compliance. In practice, reverse engineering will be harder to perform the more complex the device is, since typically no detailed design data are provided by the \ac{cots} manufacturer. In later years, guidance for use of multicore processors in \ac{aeh} have been jointly defined by many certification authorities in the Certification Authorities Software Team (CAST) paper No. 32A~\cite{CertificationAuthoritiesSoftwareTeam.}.

In practice, for \ac{cots} components different design assurance methods may be used in conjunction with each other, including the additional design assurance methods described in Appendix B in DO-254~\cite{RTCA.19.4.20001}. These are:
\begin{LaTeXdescription}

\item [Architectural mitigation] It uses architectural design features to mitigate hardware design and implementation errors. Typically, monitors and dissimilar components may be used for \ac{cots}. The safety net approach described by \cite{AIR100.02.2011} belongs to this approach.

\item [Product service experience] It identifies the service experience data and then establishes that the service experience data demonstrates that the reused functionality of the hardware is sufficiently exercised during previous uses of the hardware. \ac{easa}’s CM \cite{EASA.9.5.20121} extends DO-254’s product service experience scope with a \ac{cots} based service experience approach. The \ac{faa} report \ac{faa}\cite{Mutuel.2016} also gives guidance for the use of service experience.

\item [Advanced verification methods] It includes elemental analysis, safety-specific analysis (SSA), and formal methods. Elemental analysis requires details about the design and is therefore less suitable for \ac{cots} devices. SSA may be used as an additional design assurance approach, but may also require some details about the internal design of the device \cite{Forsberg.b}. SSA focuses on exposing design errors that could adversely affect the hardware outputs from a system safety perspective~\cite{RTCA.19.4.20001}. Formal methods apply mathematics to prove correctness of a design and has according to the authors not been extensively used for airborne \ac{cots} design assurance. The determination and validation of the usage domain described in Section 9 in \ac{easa}’s CM \cite{EASA.9.5.20121}, may also be assigned to this category. The validity of the usage domain is ensured by a set of verification activities.

\item [Extended electronic component management] It addresses control mechanisms on the \ac{cots} manufacturer by means of detailed design data, how errata is maintained and published, how device data is configuration controlled, and how change impact analysis is performed.\footnote{Extended by the means of the \ac{easa}-CM \cite{EASA.9.5.20121}.}

\item [Reverse engineering] In this way the life cycle data required by DO-254 are created from available \ac{cots} data or created from various reverse engineering techniques such as x-raying the components or detailed tests.
\end{LaTeXdescription}

These methods are used complementary as means for \ac{cots} assurance or to build up the \ac{da} strategy. \emph{The assurance process is not a proof, but rather an accumulation of elements that leads to bringing the conviction of industrial partners and certification authorities that the equipment embedding complex \ac{cots} hardware is safe}~\cite{Jean.}.  Currently, it is the \ac{cri} type of document\footnote{For \ac{faa} type certifications, issue papers are the requested type of document.}, requested by the certification authority that functions as the container for the assurance strategy of complex \ac{cots} components e.g. \acp{mpu}~\cite{Forsberg.}.

\section{Design Assurance Evaluation}\label{sec:DAE}


The previous chapter stated that a \ac{cots} assurance process is composed of different methods. The recognition of the \ac{cots} design process as a supplementary argument for the design integrity and is not often considered in industrial practise. It is required to collect as much assurance as possible, and process based evidence is one type of that.

If it could be demonstrated that the resulting integrity level of the \ac{mcu} design process is comparable to the one of DO-254; this could establish a base of trustworthiness for all other assurance activities.

The concept of \ac{da} in DO-254 is defined as \emph{planned and systematic actions used to substantiate [...], that design errors have been identified and corrected [...]}~\cite{RTCA.19.4.20001}. The extent of \ac{da} \emph{guidance}, according to the title of DO-254, goes with error mitigation strategies beyond error avoidance. Thus, it is necessary to design the hardware such that it safely performs its function as intended also under anomalous behaviour. However, the evaluation precludes these fault tolerance or fail-safe aspects in order to concentrate on the overall design approach. So activities for fault avoidance to minimize design errors are separated from fault mitigation techniques that reduce consequences triggered by these errors.



\subsection{Challenge}\label{sec:Challenge}

Aviation industry has another perspective on hardware design compared to most semiconductor manufacturer. This is logical since \ac{cots} components are targeted to domains with different claims according to design and safety goals; and the avionics industry is not among the biggest players in embedded systems. Nevertheless, avoidance of design errors is also a business case for \ac{cots} or semiconductor manufacturers in order to mitigate the risk of an economic loss by product recalls. The used approach is sufficient for this trade-off but not necessarily from a safety perspective which would cause an increased workload to reach a higher confidence. 

The possible \ac{da} misalignment is a widespread opinion in aerospace, but it could not be carefully substantiated during a literature search. Origin of this attitude can be that only the DO-254 approach and its naming is in mind when this topic is discussed. Indeed, there are also references that if the \ac{cots} design process is requirements driven it could be based on similar approach as the one described in DO-254~\cite{FAUBLADIER.20081}. Findings in~\cite{Menschner.2015} demonstrates that the verification phase of \ac{cots} components could be more comprehensive as it is required in DO-254, but because it is not compliant to DO-254 no certification credit can be claimed. Every single domain, especially safety-related one, is aware about the consequences of systematic faults and intend to reduce their likelihood of occurrence. It is generally accepted that no quantitative methods are available to measure how many possible design errors are present after a certain approach~\cite{AIR120.2006}. \emph{Because we cannot demonstrate how well we have done, we will show how hard we have tried}~\cite{Rushby.}. So specific expert judgement and best practices are promoted  for each domain as sufficient mean. These circumstance makes a comparison or evaluation quite difficult.

%

\subsection{Method}\label{sec:Method}

\subsubsection{Evaluation Goal}\label{sec:EvalGoal}

For the evaluation a terminology is needed as comparison scale between both processes. The known concept of \ac{hwdal}\footnote{On system level it is called item development assurance level.} from DO-254 is not used. This concept is not independent since it is interwoven with the avionic domain and it proportionally depends on the severity of the failure condition caused by the malfunction of the hardware. For the evaluation the naming of integrity level is proposed, since it is independent of a domain or corresponding standard. In \cite{AIR120.2006} it is stated that \ac{da} \emph{[...] establishes a level of confidence that the [...] [design] has been accomplished in a sufficiently disciplined, rigorous manner to limit the likelihood of [design] errors [...]}. The integrity level represents this confidence about the limitation of design errors as property of the \ac{da} process. 

Like a gap-analysis are the differences between the design approach of an \ac{mcu} and the DO-254 identified. The evaluation goal is to qualitatively assess these differences of the \ac{cots} design approach w.r.t. their impact of the integrity level. The evaluation focus is on avoidance of design errors compared against the DO-254 guidance. Therefore, a determined integrity level is qualitatively given in relation to the DO-254. This ensures the applicability of evaluation results for the \ac{cots} assurance case.

\subsubsection{Definition of Avionics Evaluation Base}
The DO-254 is one single document, but also has interdependencies with the ARP4754A\cite{SAEAerospace.12.2010} and DO-178C\cite{RTCA.03.131} standards on system and software level. This structure is derived by a top down development concept, starting at the aircraft system level. Safety and functional requirements are broken down to the hardware levels. This ensures the satisfaction of important safety aspects that are an overall system property. The system safety assessment is conducted in conjunction with the hardware design team, but it is not in responsibility or scope of DO-254 hardware design life cycle that started at document section 3. At this point all implications on hardware regarding safety have to be formulated as safety requirements. During the whole design life cycle the assurance of safety is still in mind, but it is formalized as requirements that have to be correctly implemented. 

The evaluation results shall be used within a \ac{da} strategy for a highly safety critical \ac{cots} component. The DO-254 introduces an assurance level approach to determine the process guidance by means of different severity stages in case of a malfunction. \ac{hwdal} A or B have to be considered for highly safety critical systems with catastrophic or hazardous failure conditions. For both, the same guidance set is applicable and thus they are considered for the evaluation. If the examination findings shall be reused for lower \ac{hwdal}, it can be adjusted afterwards.

The relevant DO-254 process guidance, considered for the evaluation, is located form chapter 3 to appendix B. For reasons of clarity and comprehensibility, further certification guidance materials for the DO-254 are skipped for the present. Despite, it is necessary to bear in mind these additional documents if a later certification is intended.  For a further evaluation iteration they will be considered. 

\subsubsection{Determination of a suitable \ac{cots} Evaluation Base}

As counterpart for the evaluation an \ac{mcu} should be selected that is potentially appropriate also in technical aspects\footnote{The technical suitability and associated certification effort are not considered here.}. Its design approach should have similarities to the DO-254 in following properties:

\begin{LaTeXdescription}
	\item[Standard Based] The \ac{mcu} manufacturer derived its design process from a widespread standard. This enables the evaluation on a higher abstraction level when comparing the DO-254 with another standard. Additionally, assessment results can be reused for other products realized in accordance to the same standard.
	\item [Process or product centred] A design process can consist of requirements on the product to define its nominal performance that shall be designed or on the process. The DO-254 does not claim requirements on the product performance. Only for \ac{da} some aspects are prescribed like architectural mitigation that have an impact on the product design.
	\item [Avoidance of systematic errors] This should be explicitly stated or be a recognizable goal of the process. An assurance level approach that defines the process rigour is a good indication for that.
\end{LaTeXdescription}

By searching for \ac{mcu} candidates on the market, automotive products developed in compliance to the ISO 26262\cite{ISO.2011} were identified. The design process is derived in compliance to the ISO 26262 which satisfies the above mentioned criteria. This process is based on a widespread standard that also focuses on hardware design. The ISO 26262 was specified for the development of functional safe electrical and/or electronic systems in a passenger car and highlighted the avoidance of design errors as one goal. This is achieved by an assurance level based design approach. Like DO-254, ISO 26262 defines process requirements to ensure the integrity of the design. The specification of nominal product performance is not considered by ISO 26262. 

Especially this third party assessed and safety aware standard compliant design process is an advantage for \ac{aeh} manufacturer. The process is still controlled by the \ac{cots} supplier but the framework is public. For a safety-aware process the reachable level of integrity is likely rather higher than for an \ac{mcu} designed for the consumer market. For several years semiconductor manufacturers provided such kind of safety \acp{mcu}. They reach a high diagnostic coverage by the integration of safety mechanisms to reduce chip external provisions. These complex \acp{mcu} are equipped with a mixture of enough computing performance and interfaces, so they are appropriate for e.g. \ac{uav} applications. 

The ISO 26262 consists of ten parts which together specify how a functional safe system, composed of software and hardware, shall be developed on vehicle level. Basically, the standard defines a top down development approach like in the aviation domain. However, a tailored life cycle is proposed, called \ac{seooc}, to enable the development of generic components, like an \ac{mcu}, independent of a specific system developement. The ISO 26262 part 5 with the title: \emph{Product development at the hardware level} and all referenced parts have to be considered for the \ac{seooc} approach and as consequence also for the evaluation.

\subsubsection{Evaluation Procedure}\label{sec:EvalProc}

DO-254 section 1.7 states that if alternative processes are used as the one described in DO-254, it has to be evaluated if objectives are fulfilled with the same level of \ac{da}. This is also the target of this evaluation, to examine if the described ISO 26262 processes enable a similar integrity level of the designed hardware. This is done for each DO-254 process expecting the certification liaison, since this is not a comparable process for the ISO 26262 as for its adherence no governmental authorities are responsible nor claim it. An accredited organization like the T\"UV S\"UD is able to verify that a semiconductor manufacturer developed an \ac{mcu} compliant to the ISO 26262. The standard do not describe the interdependencies between certification organisation and the company applying it. 

Every specified DO-254 process defines objectives and activities that formulates the guidance on how to accomplish the objectives. Associations between objectives  and activities are not explicitly described. However, for the evaluation it is helpful to understand the linkage between them, because the activities are more descriptive as the objectives, so they are more suitable for the evaluation.\footnote{Objectives and activities related to certification liaison process are omitted.} The ISO 26262 specifies for each process requirements which are formulated on a similar abstraction level as the DO-254 activities. 

The evaluation procedure starts with DO-254 activities and map them with suitable ISO 26262 requirements.\footnote{As described in \ref{sec:EvalGoal}, the evaluation or mapping is only preformed against the DO-254 processes.} In order to facilitate the mapping, the DO-254 activities are excerpted and reduced to the essentials. After this, all associations are established between DO-254 objectives, DO-254 activities and ISO 26262 requirements. In order to make a statement about the ISO 26262 compliance of each DO-254 objective, all associations form ISO 26262 requirements via DO-254 activities to DO-254 objectives are considered. From table \ref{tab:PlaProcMap} to \ref{tab:ProAProcMap} these associations are shown. To build these relations, it is necessary to interpret the equivalence between the DO-254 activities and the ISO 26262 requirements, because each standard uses its own vocabulary. For readability reasons, associated ISO 26262 requirements and DO-254 activities are only presented as item numbers. At the end of each process evaluation a qualitative statement is presented if the process according to ISO 26262 reach a comparable integrity level as the corresponding DO-254 process.

The ISO 26262 defines an assurance level approach called \ac{asil}, where  D is the level that claims the highest amount of risk reduction. All requirements and methods of the ISO 26262 applicable for \ac{asil} D are considered.

\subsection{Realisation}\label{sec:Realisation}
Before the evaluation is done according to the procedure described in \ref{sec:EvalProc}, hardware design life cycles of both standards should be compared in order to give an overview about the applied process mapping. Table \ref{tab:HWDLLPM} shows all DO-254 hardware life cycle processes relevant for the evaluation with the best associable ISO 26262 processes. Even though, the configuration management is mappable and an important process in a design life cycle, it is not considered in this evaluation since the ISO 26262 referencing another standard at this place. The reference numbering syntax of ISO 26262 processes in table \ref{tab:HWDLLPM} comprised at first a qualifier for the ISO 26262 part number, a hyphen as separator and then the chapter numbering. In section \ref{sec:Realisation} it is outlined how suitable this mapping is w.r.t. the DO-254 intention of the processes.  

\begin{table}[!t]
	\caption{Hardware Design Life Cycle Process Mapping}
	\label{tab:HWDLLPM}
	\centering
	\scriptsize
	\begin{tabularx}{.8\linewidth}{>{\hsize=.05\hsize}XX||>{\hsize=.36\hsize}XX}
		\toprule
		\multicolumn{2}{c|	|}{\bfseries DO-254} &\multicolumn{2}{c}{\bfseries ISO 26262}\\
		\bfseries Ref & \bfseries Name & \bfseries Ref & \bfseries Name \\
		\midrule
		4 & Planning Process & 5-5 & Initiation of Product Development at the Hardware Level\\\midrule
		5.1 & Requirements Capture Process & 5-6\par 8-6 & Specification of Hardware Safety Requirements\\\midrule
		5.2 & Conceptual Design Process & 5-7.4.1 & Hardware Architectural Design\\\midrule
		5.3 \par 5.4 & Detailed Design and \par Implementation Process & 5-7.4.2 & Hardware Detailed Design\\\midrule
		5.5 & Product Transition Process & 5-7.4.5&Product, Operation and Decommissioning\\\midrule
		6 & Validation and Verification Process&8-9\par 8-6\par 5-6\par 5-7.4.4\par 5-10 &  Verification \\\midrule
		
	7 & Configuration Management Process&8-7\par 8-8 & Configuration and Change Management\\\midrule

		8 & Process Assurance & 2-6 & Safety Management during the Concept and the Product Development \\
		
		\bottomrule
	\end{tabularx}
	
\end{table}

Neither ISO 26262 nor DO-254 prescribe a life cycle model, since this is not a part of the evaluation.\footnote{The mentioned V-model in ISO 26262 is just for reference.} Both standards define a reference process model that indicates the idealized process sequencing and remarks that they may be iterative entered and re-entered with feedback flows. This depends on many factors and has to be defined by the applicant.

\subsubsection{Planning Process}

The value of planning the hardware design life cycle are considered by both standards in the same way. An \ac{mcu} developed compliant to ISO 26262 is realized as \ac{seooc} and has to include planning activities from ISO 26262 part 2 at system level to fulfil the planning needs as a whole.
%

The DO-254 objective one is fully satisfied by the defined activities in ISO 26262, see mapping in table \ref{tab:PlaProcMap}. Subcontractor dependencies and the control of stipulated activities are extensively described in ISO 26262, since this is a widespread approach in the automotive industry. The dedicated request for design standards is not part of the planning phase of the ISO 26262, but it is mentioned during the design phase. The diligence on definition and description of the hardware verification and design environment are  on an equivalent level. Tool assessment is addressed by ISO 26262 as an explicit supporting process. Like in the DO-254, it should be ensure that used tools for design and verification perform on an acceptable level of confidence. Although, this is a relevant aspect for avoiding design errors, it is not addressed in this evaluation in detail. 

\begin{table}[!t]
	\caption{Planning Process Mapping}
	\label{tab:PlaProcMap}
	\centering
	\scriptsize
	
		\begin{tabularx}{\linewidth}{>{\hsize=.2\hsize}X>{\hsize=.05\hsize}X||X}
		\toprule
		\multicolumn{2}{l}{\bfseries DO-254 Objectives} &\multicolumn{1}{c}{}\\
		\multicolumn{3}{p{8cm}}{\begin{enumerate}

				\item \emph{The hardware design life cycle processes are defined. }
				\item \emph{Standards are selected and defined. }
				\item \emph{The hardware development and verification environments are selected or defined.}

			\end{enumerate}
		}\\

	\end{tabularx}

		\begin{tabularx}{0.6\linewidth}{>{\hsize=0.5cm}X>{\hsize=.5cm}X||>{\hsize=2cm}l}
	
	\midrule
	\multicolumn{2}{c||}{\bfseries DO-254} &\multicolumn{1}{c}{\bfseries ISO 26262}\\
		\midrule
		\bfseries Obj \par Nr &\bfseries Act Nr & \bfseries Req Nr \\
		\midrule
		1. & 1. &5-5.4.1; 2-6.4.3.5e; 2-6.4.4.1\\
		\midrule
		1. &4. & 2-6.4.3.5g,h\\
		\midrule
		1. &5. &  2-6.4.3.5g,h \\
		\midrule
		1. &9. & 2-6.4.3.5f \\		
		\midrule
		2. &3. & 5-7.4.2.4\footnote{Origin outside the mapped process.} \\

		\midrule
		3. &6.  & 5-5.4.1; 2-6.4.3.5l  \\
		\midrule
3. &8. &2-6.4.4.2; 2-7.4.1\\		
		\midrule
3. &10. & 2-6.4.8.1; 2-6.4.9.2 ;2-6.4.7.1 \\
		\midrule
3. &11.  & 2-6.4.7.1  \\

		\bottomrule
	\end{tabularx}
\end{table}

	\begin{LaTeXdescription} 
	\item[Evaluation findings] The planning process of the ISO 26262 is broadly the same, see mapping in table \ref{tab:PlaProcMap}. No explicit need for creation of own defined hardware design standards is stated in the planning process, but this difference is not enough to deprecate the possible integrity level. It is assumed, that such kind of \emph{coding guidelines} are part of a robust design flow and claimed by ISO 26262 in the hardware detailed design process. The integrity level is evaluated as equal $\rightarrow$.\footnote{Tool assessment approach is not considered.}
\end{LaTeXdescription}

\subsubsection{Requirements Capture Process}
Both standards describe a requirement driven design flow and define a dedicated process for the capturing and specification of requirements. The mapping in table \ref{tab:ReqProcMap} shows that six DO-254 activities are associated to objective one. They describe that requirements from different sources shall be reasonably identified, documented and linked to each other. Safety and derived requirements are emphasized. The source of a requirement could be the system level process or design constraints. A derived requirement is a requirement that is an interpretation of an allocated system requirement or a consequence of a design constraint and is defined to be implementable on the hardware level. The ISO 26262 also knows concepts of derived requirements but concentrates on safety related requirements. Requirements without safety relationships can be managed outside of the ISO 26262 life cycle. This is not the case for the DO-254. Bidirectional traceability between hierarchical requirements levels are requested by both standards. According to DO-254 objective two and three, the ISO 26262 specifies that all hardware requirements have to be verified with higher level specifications.

\begin{table}[!t]
	\caption{Requirement Capture Process Mapping}
	\label{tab:ReqProcMap}
	\centering
	\scriptsize
	
		\begin{tabularx}{\linewidth}{>{\hsize=.2\hsize}X>{\hsize=.05\hsize}X||X}
		\toprule
		\multicolumn{2}{l}{\bfseries DO-254 Objectives} &\multicolumn{1}{c}{}\\
		\multicolumn{3}{p{8cm}}{\begin{enumerate}

				\item \emph{Requirements are identified, defined and documented. [...] }
				\item \emph{Derived requirements produced are fed back to the appropriate process. }
				\item \emph{Requirements omissions and errors are provided to the appropriate process for resolution.}

			\end{enumerate}
		}\\

	\end{tabularx}
	
		\begin{tabularx}{0.5\linewidth}{>{\hsize=0.5cm}X>{\hsize=.5cm}X||>{\hsize=2cm}l}
	
	\midrule
	\multicolumn{2}{c||}{\bfseries DO-254} &\multicolumn{1}{c}{\bfseries ISO 26262}\\
\midrule
\bfseries Obj \par Nr &\bfseries Act Nr & \bfseries Req Nr \\
		\midrule
		1. & 1. & 8-6.4.2.3; 5-6.4.2\\
		\midrule
		1. &2. & 8-6.4.2.5a,c\\
		\midrule
		1. &3. &  5-6.4.1 \\
		\midrule
		1. &4. & 5-6.4.1; 5-6.4.2  \\
		\midrule
		1. &6. & 8-6.4.2.4e  \\
		\midrule
		1. &8. & 8-6.4.3.1; 8-6.4.3.2a,b  \\
		\midrule
		
		2. &5. & 5-6.4.9  \\
		\midrule
		3. &7. & 5-6.4.9  \\
		
		\bottomrule
	\end{tabularx}
	
\end{table}

\begin{LaTeXdescription}
	\item[Evaluation findings] The concentration on safety-related requirements, in the requirements capture phase of the ISO 26262, could cause misunderstandings by managing non-safety-related requirements in another way. However, according to ISO 26262 design process all requirements, even non-safety requirements, have to be handled within one development process. The DO-254 clearly states that all allocated requirements have to be considered. It is a debatable point whether it is practical in a real development to separate the managing of some requirements, but it is possible.  Such a situation could increase the potential of design errors. This cause an integrity level deprecation of the ISO 26262 $\searrow$.
\end{LaTeXdescription}

\subsubsection{Conceptual Design Process}

Within the conceptual design phase, the hardware architecture shall be defined. All components are described with their interactions with one another on a higher level description, like the detailed design afterwards. 


The ISO 26262 is also aligned with the objective number one, refer table \ref{tab:ConDesProcMap}. The requirements implemented in the hardware architecture shall be traceable to the lowest level of hardware component. If safety specific architectural constraints are formulated as hardware safety requirements, they are considered in the architecture. Unused functions are addressed in ISO 26262 if they have a safety impact. In this case, their independence has to be demonstrated. The feedback of derived requirements according DO-254 objective two is not explicitly described in ISO 26262. Problems with requirements are communicated in the verification phase as specified in ISO 26262.

\begin{table}[!t]
	\begin{minipage}[t]{\linewidth}
	\caption{Conceptual Design Process Mapping}
	\label{tab:ConDesProcMap}
	\centering
	\scriptsize
		\begin{tabularx}{\linewidth}{>{\hsize=.2\hsize}X>{\hsize=.05\hsize}X||X}
		\toprule
		\multicolumn{2}{l}{\bfseries DO-254 Objectives} &\multicolumn{1}{c}{}\\
		\multicolumn{3}{p{8cm}}{\begin{enumerate}
				
					\item \emph{The hardware item conceptual design is developed consistent with its requirements. }
				\item \emph{Derived requirements produced are fed back to the requirements capture or other appropriate processes. }
				\item \emph{Requirements omissions and errors are provided to the appropriate process for resolution.}

			\end{enumerate}
		}\\

	\end{tabularx}
	
		\begin{tabularx}{0.6\linewidth}{>{\hsize=0.5cm}X>{\hsize=.5cm}X||>{\hsize=2cm}l}
	
	\midrule
	\multicolumn{2}{c||}{\bfseries DO-254} &\multicolumn{1}{c}{\bfseries ISO 26262}\\
\midrule
\bfseries Obj \par Nr &\bfseries Act Nr & \bfseries Req Nr \\
		\midrule
		1. & 1. &5-7.4.1.1\\
		\midrule
		1. &2. & 5-7.4.1.1; 5-7.4.1.5; 5-7.4.3.5\footnote{\label{fot:1}Origin outside the mapped process.}\\
		\midrule
		1. &5. &  5-7.4.1.1; 5-7.4.1.6 \\
		\midrule
		2. &3. &   --\\
		\midrule
		3. &4. & 5-7.4.4.2\footref{fot:1} \\
		
		\bottomrule
	\end{tabularx}
\end{minipage}
\end{table}

\begin{LaTeXdescription}
	\item[Evaluation findings] The ISO 26262 fulfils the objectives, see table \ref{tab:ConDesProcMap}. The DO-254 emphasizes safety-related aspects of the hardware architecture, that should usually be introduced by requirements coming from the safety assessment. ISO 26262 does not explicit mention that also in this phase requirements could be derived, but the incremental re-entering of the requirement specification phase could be performed\footnote{See footnote table \ref{tab:ConDesProcMap}.}. The ISO 26262 process is considered on an equal level of integrity $\rightarrow$.
\end{LaTeXdescription}

\subsubsection{Detailed Design and Implementation Process}

Detailed design and implementation are separate processes in DO-254. ISO 26262 does not describe a dedicated phase for implementation of the design. It is assumed as a part of the overall hardware design phase, that is ending with a component ready for test activities. Both tables \ref{tab:DetDesProcMap} and \ref{tab:ImplProcMap} show that most of the mapped ISO 26262 requirements do not originating from the dedicated ISO 26262 processes used as evaluation counterpart. So the concrete activities mentioned by the DO-254 processes have to be associated, on the ISO 26262 side, with requirements from other phases.

The first objective of the detailed design process is respected by the ISO 26262. However, the highlighting of safety awareness during this phase is not done by ISO 26262 since these features have to be defined as hardware safety requirements. One obvious difference between both standards is that the DO-254 claims requirement traceability down to the detailed design, whereas for the ISO 26262 it ends at the conceptual design. 

The objectives related to derived requirements and omissions in both DO-254 processes\footnote{See table \ref{tab:DetDesProcMap} and \ref{tab:ImplProcMap}.} are treated by the ISO 26262 in the same way. There is no certain attention given on this aspect, expect that the hardware has to be designed in compliance to its requirements. 

For the implementation process objective one and two, the ISO 26262 do not prescribe any requirements that could be directly traced. It is implicitly included in the design phase.

\begin{table}[!t]
		\begin{minipage}[t]{\linewidth}
	\caption{Detailed Design Process Mapping}
	\label{tab:DetDesProcMap}
	\centering
	\scriptsize
	\begin{tabularx}{\linewidth}{>{\hsize=.2\hsize}X>{\hsize=.05\hsize}X||X}
		\toprule
		\multicolumn{2}{l}{\bfseries DO-254 Objectives} &\multicolumn{1}{c}{}\\
		\multicolumn{3}{p{8cm}}{\begin{enumerate}

				\item \emph{The detailed design is developed from the hardware item requirements and conceptual design data.  }
				\item \emph{Derived requirements are fed back to the conceptual design process or other appropriate processes. }
				\item \emph{Requirements omissions and errors are provided to the appropriate process for resolution.}
			\end{enumerate}
		}\\

	\end{tabularx}
	
	\begin{tabularx}{0.4\linewidth}{>{\hsize=0.5cm}X>{\hsize=.5cm}X||>{\hsize=2cm}l}
	
	\midrule
	\multicolumn{2}{c||}{\bfseries DO-254} &\multicolumn{1}{c}{\bfseries ISO 26262}\\
	\midrule
	\bfseries Obj \par Nr &\bfseries Act Nr & \bfseries Req Nr \\
		\midrule
		1. & 1. &5-7.4.4.1\footnote{\label{fot:2}Origin outside the mapped process.}\\
		\midrule
		1. &2. & 5-7.4.1.1\footref{fot:2}\\
		\midrule
		1. &3. &5-7.4.1.6\footref{fot:2}\\
				\midrule
		1. &4. &  5-7.4.3.5\footref{fot:2}\\
				\midrule
		1. &5. &  5-7.4.2.2 \\
		\midrule
		2. &6. &  -- \\
		\midrule
		3. &7. & 5-7.4.4.2 \footref{fot:2} \\
		
		\bottomrule
	\end{tabularx}
\end{minipage}
\end{table}

\begin{table}[!t]
	\begin{minipage}[t]{\linewidth}
		\caption{Implementation Process Mapping}
		\label{tab:ImplProcMap}
		\centering
		\scriptsize
		
			\begin{tabularx}{\linewidth}{>{\hsize=.2\hsize}X>{\hsize=.05\hsize}X||X}
			\toprule
			\multicolumn{2}{l}{\bfseries DO-254 Objectives} &\multicolumn{1}{c}{}\\
			\multicolumn{3}{p{8cm}}{\begin{enumerate}
					
						\item \emph{A hardware item is produced which implements the hardware detailed design using representative manufacturing processes.}
					\item \emph{The hardware item implementation, assembly and installation data is complete.}
					\item \emph{Derived requirements are fed back to the detailed design process or other appropriate processes.}
					\item \emph{Requirements omissions and errors are provided to the appropriate process for resolution.}

				\end{enumerate}
			}\\

		\end{tabularx}
		
		\begin{tabularx}{0.40\linewidth}{>{\hsize=0.5cm}X>{\hsize=.5cm}X||>{\hsize=2cm}l}
	
	\midrule
	\multicolumn{2}{c||}{\bfseries DO-254} &\multicolumn{1}{c}{\bfseries ISO 26262}\\
\midrule
\bfseries Obj \par Nr &\bfseries Act Nr & \bfseries Req Nr \\
			\midrule
			1. & 1. &--\\
			\midrule
			2. &1. &  -- \\
			\midrule
			3. &2. &  --\\
			\midrule
			4. &3. &  5-7.4.4.2\footnote{Origin outside the mapped process.} \\
			
			\bottomrule
		\end{tabularx}
	\end{minipage}
\end{table}

\begin{LaTeXdescription}
	\item[Evaluation findings] The DO-254 separates the design from the implementation in a better way. The ISO 26262 gives no hint to start the implementation with the same means the later production product is produced. This ensures that most of the verification activities can be performed with a hardware really close to the final versions. This deviation and the fact that the requirement traceability is not continuously established down to the detailed design, leads to the implication that the ISO 26262 has a lower integrity level w.r.t. these phases $\searrow$.
\end{LaTeXdescription}

\subsubsection{Product Transition Process}

The series production is out of the scope of the DO-254, but preparation of data for production is treated. As shown in table \ref{tab:ProTranProcMap}, the mapped ISO 26262 requirements originate from the system level production and hardware level phase. Together they fulfil the DO-254 objectives one, two and four. The emphasized feedback of derived requirements claimed by objective three could not be determined in ISO 26262 like for the other processes. 



\begin{table}[!t]
	\begin{minipage}[t]{\linewidth}
		\caption{Product Transition Process Mapping}
		\label{tab:ProTranProcMap}
		\centering
		\scriptsize
		
			\begin{tabularx}{\linewidth}{>{\hsize=.2\hsize}X>{\hsize=.05\hsize}X||X}
			\toprule
			\multicolumn{2}{l}{\bfseries DO-254 Objectives} &\multicolumn{1}{c}{}\\
			\multicolumn{3}{p{8cm}}{\begin{enumerate}

					\item \emph{A baseline is established that includes all design and manufacturing data needed to support the consistent replication of the hardware item.}
					\item \emph{Manufacturing requirements related to safety are identified and documented and manufacturing controls are established.}
					\item \emph{Derived requirements are fed back to the implementation process or other appropriate processes.}
					\item \emph{Errors and omissions are provided to the appropriate processes for resolution.}
					
				\end{enumerate}
			}\\

		\end{tabularx}
		\begin{tabularx}{0.48\linewidth}{>{\hsize=0.5cm}X>{\hsize=.5cm}X||>{\hsize=2cm}l}
	
	\midrule
	\multicolumn{2}{c||}{\bfseries DO-254} &\multicolumn{1}{c}{\bfseries ISO 26262}\\
\midrule
\bfseries Obj \par Nr &\bfseries Act Nr & \bfseries Req Nr \\
			\midrule
			1. & 1. &4-11.4.2.2\footnote{\label{fot:6}Origin outside the mapped process.}\\
			\midrule
			1. &2. & 4-11.4.1.1\footref{fot:6}\\
			\midrule
			1. &5. &4-11.4.1.1\footref{fot:6}\\
			\midrule
			2. &4. &5-7.4.5.1; 5-7.4.5.2\\
			\midrule
			3. &3. &--  \\
			\midrule
			4. &6. &  4-11.4.2.3\footref{fot:6} \\
			
			\bottomrule
		\end{tabularx}
	\end{minipage}
\end{table}

\begin{LaTeXdescription}
	\item[Evaluation findings] The ISO 26262 plans in a very similar manner the product transition like DO-254. Derived requirements feedback and the concrete manufacturer control are not mentioned by ISO 26262, but this does not lead to a deprecation for the evaluation. The resulted integrity level of this process is comparable $\rightarrow$.
\end{LaTeXdescription}

\subsubsection{Validation and Verification Process}
In DO-254 and ISO 26262, validation and verification are supporting processes. This means, that they are concurrently performed with the former evaluated design processes. Between validation and verification no differentiation is stated in ISO 26262 in the way like in the DO-254. The supporting process verification in ISO 26262 part 8 is the conceptual root where the process is described in general. At hardware level, validation tasks, in the sense of DO-254, are linked at the requirement specification\footnote{Section 5-6 in\cite{ISO.2011}.}. Further, verification according to DO-254 is associated to hardware design verification\footnote{Section 5-7.4.4 in\cite{ISO.2011}.} and testing\footnote{Section 5-10 in\cite{ISO.2011}.} on hardware level. This is the reason for the variation in ISO 26262 references in table \ref{tab:ValProcMap} and \ref{tab:VerProcMap}. The DO-254 also describes validation and verification tasks at hardware level, but focused these aspects in the same document chapters as the process description. Hence, the ISO 26262 is applied to the entire system development it defines another document structure.

Table \ref{tab:ValProcMap} shows that for every described DO-254 activity one or more ISO 26262 requirements could be mapped to adequately fulfil the DO-254 objectives. The validation or the verification of lower level hardware requirements against the system specification is claimed in a diligent manner. The ISO 26262 postulates validation means like inspections, reviews and semi-formal methods.

Also, table \ref{tab:VerProcMap} demonstrates a sufficient mapping of DO-254 activities with ISO 26262 requirements of comparable content. The DO-254 stated that hardware requirements can be verified at a higher hierarchical level. It is not necessary to verify all requirements at the level of the detailed design. However, the traceability down to that level should be established to verify the completeness or test coverage. The ISO 26262 verification means do not induce the same rigour as claimed by the advanced verification means in DO-254 appendix B. The level of independence for verification, requested in DO-254 appendix A, is not considered in ISO26262\footnote{Independence is only requested for confirmation measures, see section \ref{sec:procAs}.}.






\begin{table}[!t]
	\begin{minipage}[t]{\linewidth}
		\caption{Validation Process Mapping}
		\label{tab:ValProcMap}
		\centering
		\scriptsize
		
			\begin{tabularx}{\linewidth}{>{\hsize=.2\hsize}X>{\hsize=.05\hsize}X||X}
			\toprule
			\multicolumn{2}{l}{\bfseries DO-254 Objectives} &\multicolumn{1}{c}{}\\
			\multicolumn{3}{p{8cm}}{\begin{enumerate}

					\item \emph{Derived hardware requirements against which the hardware item is to be verified are correct and complete.}
					\item \emph{Derived requirements are evaluated for impact on safety.}
					\item \emph{Omissions and errors are fed back to the appropriate processes for resolution.}

				\end{enumerate}
			}\\

		\end{tabularx}

		\begin{tabularx}{0.62\linewidth}{>{\hsize=0.5cm}X>{\hsize=.5cm}X||>{\hsize=2cm}l}
	
	\midrule
	\multicolumn{2}{c||}{\bfseries DO-254} &\multicolumn{1}{c}{\bfseries ISO 26262}\\
\midrule
\bfseries Obj \par Nr &\bfseries Act Nr & \bfseries Req Nr \\
			\midrule
			1. & 1. &5-6.4.2; 8-6.4.3.4\\
			\midrule
			1. &2. &8-6.4.3.1e; 8-6.4.3.3; 8-9.4.1.1c \\
			\midrule
			1. &4. &5-6.4.9b\\
						\midrule
			1. &5. &5-6.4.9d\\
						\midrule
			1. &6. &8-9.4.3.2a\\
			\midrule
			2. &3. &5-6.4.2\\
			\midrule
			3. &7. & 5-7.4.4.2\\

			\bottomrule
		\end{tabularx}
	\end{minipage}
\end{table}

\begin{table}[!t]
	\begin{minipage}[t]{\linewidth}
		\caption{Verification Process Mapping}
		\label{tab:VerProcMap}
		\centering
		\scriptsize
		
			\begin{tabularx}{\linewidth}{>{\hsize=.2\hsize}X>{\hsize=.05\hsize}X||X}
			\toprule
			\multicolumn{2}{l}{\bfseries DO-254 Objectives} &\multicolumn{1}{c}{}\\
			\multicolumn{3}{p{8cm}}{\begin{enumerate}

								\item \emph{Derived hardware requirements against which the hardware item is to be verified are correct and complete.}
					\item \emph{Derived requirements are evaluated for impact on safety.}
					\item \emph{Omissions and errors are fed back to the appropriate processes for resolution.}
					
				\end{enumerate}
			}\\

		\end{tabularx}
	
\begin{tabularx}{0.6\linewidth}{>{\hsize=0.5cm}X>{\hsize=.5cm}X||>{\hsize=2cm}l}
	
	\midrule
	\multicolumn{2}{c||}{\bfseries DO-254} &\multicolumn{1}{c}{\bfseries ISO 26262}\\
			\midrule
			\bfseries Obj \par Nr &\bfseries Act Nr & \bfseries Req Nr \\
			\midrule
			1. & 1. &5-10.4.5\\
			\midrule
			1. &2. &8-9.4.2.3; 8-9.4.2.1; 8-9.4.1.1b \\
			\midrule
			2. &3. &8-9.4.3.2a\\
			\midrule
			2. &5. &8-9.4.3.2\\
			\midrule
			3. &4. &8-9.4.1.1\\
			\midrule
			4. &6. &5-7.4.4.2\\

			\bottomrule
		\end{tabularx}
	\end{minipage}
\end{table}

\begin{LaTeXdescription}
	\item[Evaluation findings] Both standards have different meanings of the term validation and verification. This is not a show stopper for reaching the same level of integrity. However, the different depth of traceability, the omission of independence and the less rigour for verification means in ISO 26262 results in a lower integrity level as requested by DO-254 $\searrow$ .
\end{LaTeXdescription}

\subsubsection{Process Assurance}\label{sec:procAs}
 In contrast to the DO-254, the ISO 26262 prescribes the organizational structure for the process assurance activities in form of responsibilities and their administrations rights. The requested independence by DO-254 is satisfied by ISO 26262. Confirmation measures are performed by persons from a different department or organization. The DO-254 objectives are fulfilled by a small set of ISO 26262 requirements. The reason is, that two kind of confirmation measures are responsible to verify work products, output of processes, and the adherence of the plans. Table \ref{tab:ProAProcMap} shows that all DO-254 activities could be mapped to ISO 26262 requirements with a similar intention.

\begin{table}[!t]
	\begin{minipage}[t]{\linewidth}
		\caption{Process Assurance Mapping}
		\label{tab:ProAProcMap}
		\centering
		\scriptsize
		
			\begin{tabularx}{\linewidth}{>{\hsize=.2\hsize}X>{\hsize=.05\hsize}X||X}
			\toprule
			\multicolumn{2}{l}{\bfseries DO-254 Objectives} &\multicolumn{1}{c}{}\\
			\multicolumn{3}{p{8cm}}{\begin{enumerate}

						\item \emph{Life cycle processes comply with the approved plans.}
					\item \emph{Hardware design life cycle data produced complies with the approved plans.}
					\item \emph{The hardware item used for conformance assessment is built to comply with the associated life cycle data. }

				\end{enumerate}
			}\\

		\end{tabularx}
		
		\begin{tabularx}{0.48\linewidth}{>{\hsize=0.5cm}X>{\hsize=.5cm}X||>{\hsize=2cm}l}
	
	\midrule
	\multicolumn{2}{c||}{\bfseries DO-254} &\multicolumn{1}{c}{\bfseries ISO 26262}\\
\midrule
\bfseries Obj \par Nr &\bfseries Act Nr & \bfseries Req Nr \\
			\midrule
			1. & 2. &2-6.4.3.2; 2-6.4.7.1\\
			\midrule
			1. &3. & 2-6.4.8.1\\
			\midrule
			1. &4. &2-6.4.7.1; 2-6.4.8.1\\
						\midrule
			1. &6. &2-6.4.8.1\\
						\midrule
			1. &7. &2-6.4.8.1\\
			\midrule
			2. &1. &2-6.4.7.1\\
			\midrule
			3. &5. & 2-6.4.7.1 \\

			\bottomrule
		\end{tabularx}
	\end{minipage}
\end{table}

\begin{LaTeXdescription}
	\item[Evaluation findings]  The required degree of independence during process assurance is satisfied by ISO 26262. The DO-254 objectives are addressed with an appraoch that focused on the same kind of activities. Hence, for this process the ISO 26262 contributes to the product integrity on the same level as the DO-254 $\rightarrow$.
\end{LaTeXdescription}

\section{Conclusion}\label{sec:Conclusion}
The performed evaluation could be considered as a first step to get an impression about the differences between the both standards in avoidance of hardware design errors. The applied qualitative measure of integrity level, in relation to the DO-254, results in a tendency that the ISO 26262 will not reach the necessary design assurance requested by the DO-254. This is derived from the evaluation findings in section \ref{sec:Realisation}, where discrepancies for following processes were determined: requirements capture, detailed design, implementation, validation and verification process.

An evaluation on the abstract level of standard documents is not sufficient to come to a final result, if an \ac{amcu} developed with an ISO 26262 compliant process has a lower integrity level as requested by DO-254. The provided evaluation findings can be reused for a deeper analysis on the interpreted level of the standard. This means, to evaluate an ISO 26262 compliant development process of an \ac{mcu} manufacturer with the additional certification authority guidance for DO-254, that give a better clarity on interpretation of the DO-254 document. For this purpose, a contact is established to a \ac{cots} manufacturer. 

As a further part of this deeper evaluation, the ISO 26262 certification, authorised by an independent accredited organisation like T\"UV S\"UD, is considered. It is necessary to understand how such a certificate is issued. What are the inspection methods and the focused areas? These have to be compared against the certification authority approach. Findings of that evaluation are especially valuable, because it is supposed that \ac{mcu} manufacturers are willing to disclose same documents with a certification authority. 

In view of ever growing market share of hardware components for safety-related embedded systems~\cite{AndrewHopkins.09.2016}, such evaluations findings will become of more relevance. It is a benefit for certification authorities to understand the meaning of such \ac{cots} component standards w.r.t. to their own regulations and accepted means of compliance. Avionic and semiconductor industries could both profit from products that consider aviation concerns.

\section*{Acknowledgment}
This paper is sponsored by the Airbus Defense and Space endowed professorship "System Technology for safety-related Applications" supported by "Stifterverband f\"{u}r die Deutsche Wissenschaft e.V.".

MDH’s work in this paper is supported by the Swedish Knowledge Foundation within the project DPAC.

\bibliographystyle{IEEEtran}

\bibliography{Literatur}


\acuse{cri}

\end{document}